# 3D Real-Time Supercomputer Monitoring

Bill Bergeron, Matthew Hubbell, Dylan Sequeira, Winter Williams,
William Arcand, David Bestor, Chansup, Byun, Vijay Gadepally, Michael Houle, Michael Jones,
Anna Klien, Peter Michaleas, Lauren Milechin, Julie Mullen Andrew Prout, Albert Reuther,
Antonio Rosa, Siddharth Samsi, Charles Yee, Jeremy Kepner
MIT

*Abstract*—Supercomputers are complex systems producing vast quantities of performance data from multiple sources and of varying types. Performance data from each of the thousands of nodes in a supercomputer tracks multiple forms of storage, memory, networks, processors, and accelerators. Optimization of application performance is critical for cost effective usage of a supercomputer and requires efficient methods for effectively viewing performance data. The combination of supercomputing analytics and 3D gaming visualization enables real-time processing and visual data display of massive amounts of information that humans can process quickly with little training. Our system fully utilizes the capabilities of modern 3D gaming environments to create novel representations of computing hardware which intuitively represent the physical attributes of the supercomputer while displaying real-time alerts and component utilization. This system allows operators to quickly assess how the supercomputer is being used, gives users visibility into the resources they are consuming, and provides instructors new ways to interactively teach the computing architecture concepts necessary for efficient computing.

## I. Introduction

Optimizing the usage of and effectiveness in supercomputing is directly related to having exceptional Data Center Infrastructure Management (DCIM) tools. As published in 2015 [1] and 2019 [2] the Lincoln Laboratory Supercomputing Center (LLSC) has been developing MM3D which utilizes High Performance Computing (HPC) analytics and the Unity 3D gaming platform to process and display the massive amounts of data produced in near real time by our systems. Using our internally developed Dynamic Distributed Dimensional Data Model (D4M) [3] [4] to digest and process data from over 3,000 HPC components and 6,366 environmental sensors totalling 80 million data points per day, we transform this raw Big Data into actionable information. The 3D gaming environment is then used to take the manageable, though still considerable, data set and visualize it quickly in an intuitive manner. Our goal is achieving situational awareness for the many system support staff, systems administrators, managers, and end users on our HPC systems to facilitate the quick resolution of system problems often caused by the inefficient or improper utilization of HPC resources.

## II. Current State of System Monitoring

System monitoring of large supercomputing environments tends to focus on data collection and rely on enterprise monitoring tools, difficult to configure open source tools with limited support, or expert analysts to interpret the data. This leads to an inefficient work flow as hardware, system configuration, or poor system optimization problems are not realized in real time, if at all. Problem resolution then requires engaging experts, typically with root level system access, to interpret the data and logs to troubleshoot the system. Solving this work flow issue is a point of focus on MIT SuperCloud systems. Our HPC systems specialize in interactive supercomputing, allowing for multiple users to quickly run on the same hardware simultaneously. This makes proper monitoring and visualization of the system status critical.

### A. Data Collection and Data Deluge

The focus of most infrastructure management systems is collection, and in some cases aggregation, of system and sensor time-series data. Operating systems and firmware generate event logs tracking system conditions, warnings, and alerts. HPC Storage devices continuously generate statistics specific to their operation and networking hardware generates similar operational information and event logs. Job schedulers, which typically manage the resource distribution on an HPC system, collect job data tracking the resources requested and work performed by the users on the system. Additionally, data centers closely monitor cooling and environmental systems that regulate the buildings or enclosures housing the HPC systems. At the LLSC we typically gather approximately 13 million processed data points a day and an order of magnitude more in raw data. The sum total of the available data should allow for full situational awareness but the data volume and variety can be a challenge. There is a vast difference between having the data available and having the data accessible in a timely and effective manner.

### B. DCIM Tools

Existing DCIM tools such as Collectd [5], Telegraf [6], Ovis [7], and InfluxDB [8] are entirely text-based and require users to adhere to strict functions and syntax. Users enter SQL-style queries to the database and receive responses in array format. An example of an InfluxDB query is listed below. In this case, a HPC administrator is attempting to diagnose a central storage slowdown by assessing whether any of the currently running user workflows exceed a reasonable amount of metadata server load as measured by the number of files opened in the past 10 minutes:

```
> SELECT "jobid", ROUND(TOP("opens", 10)) AS "opens_last_10m"
  FROM (SELECT NON_NEGATIVE_DERIVATIVE(MEAN("jobstats_open"), 10m) AS "opens"
  FROM "lustre" WHERE time >= now() - 10m AND time <= now() GROUP BY jobid, time(10m))

   name: lustre
   time                    jobid          opens_last_10m
   ----                    -----          --------------
   2021-06-10T15:40:00Z    23159087       893817
   2021-06-10T15:40:00Z    23159084       496977
   2021-06-10T15:40:00Z    23184225       202221
   2021-06-10T15:40:00Z    23184272       201494
   2021-06-10T15:40:00Z    23184274       200337
   2021-06-10T15:40:00Z    23184271       199973
```

The above example illustrates various challenges in existing infrastructure management tools. Queries require significant specificity; the above simple example required a nested sub-query employing 5 data transformation functions to return 10 data records. This query/response structure requires the user to either know exactly what they are looking for or make multiple searches. Additionally, the syntax contains 163 characters and a single typographical error would result in a failure message rather than the intended result. This reduces accessibility for novice users and slows response time to issues. Finally, the output is not displayed contextually, and in general humans are significantly better at recalling pictures than words [9]. A text-based display makes it difficult for HPC system administrators to understand a large volume of data quickly.

Current DCIM tools are designed to answer specific questions but do not give the user a broader understanding of the system as a whole. This forces a reactive approach to data center management since HPC system administrators can only view a few lines of data at a time, encouraging a less proactive tendency to wait for a component of the system to degrade or break before they are able to identify and correct issues with problematic or inefficient user workflows.

### C. Visualization Tools

Graph-based visualization tools such as Grafana [10], Nagios [11], and Datadog [12] are designed around the use of custom dashboards for the presentation of DCIM data. An example of such a dashboard is shown in Figure 1.

These dashboards show that you can employ visual methods to fit large amounts of information on the screen at one time. This approach does require significant expertise to quickly interpret; users can only zoom in on one graph at a time, which means that they still have to know what they are looking for in advance. The 2D graphs do not provide an intuitive association between the data and what the data is representing.

The overall health of a supercomputer cannot easily be gauged with a quick glance at Grafana or Nagios dashboard. The same color or shape could mean different things on two dashboards. Users can employ customization options but lose the ability to easily communicate with teammates and outside organizations.

3D visualization tools, such as the proprietary ones offered by HP, Siemens, and Sunbird [13], offer isometric views of data centers and models of individual nodes. These tools represent an initial step towards making use of the 3D medium, but do not harness the full power of immersive 3D environments pioneered by the gaming industry.

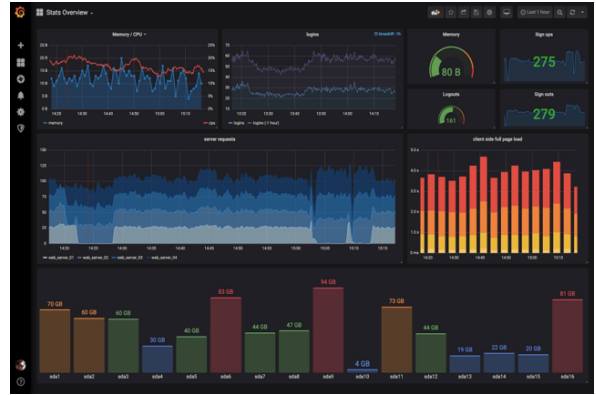

Fig. 1: *Grafana [10] visualization tools widely used for Data Center Infrastructure Management (DCIM)*

### D. Human Analysis

Despite the widespread availability of high-level visualization and monitoring tools, the primary method for real-time system troubleshooting remains manual intervention by trained experts employing elevated system privileges, typically involving logging into multiple system components and manually running commands, checking logs, and evaluating outputs.

An experienced HPC system administrator, with elevated system privileges, will go through a number of steps to determine the root cause of a problem using similar information that is also gathered elsewhere. Some of the most common and troublesome systemic issues on a supercomputer occur with the central storage, which is presented as a monolithic multi-petabyte entity but consists of hundreds of hard drives managed by a collection of individual Metadata servers (MDS) and Object Store servers (OSS).

An HPC admin, without sophisticated monitoring tools, would begin by determining the nature of the problem once alerted of an issue. This painstaking and repetitive process comprises a number of steps: determining the overall health of the system at a hardware level by interfacing with each affected component as well as verifying failover/redundancy status and assessing whether any recent configuration changes could have resulted in the behavior being exhibited. Next, individually probing each of many storage servers in an attempt to isolate the client or cluster of client nodes causing the problematic over consumption of resources, and then, if a pattern is identified, connecting to each of these client nodes to verify that the offending user workload has been found. Each of these steps typically involves searching through log files on affected systems to identify common issues.

### III. APPROACH

The approach used by MM3D to manage the data overload is twofold: first, to generate a reliable, structured, and adaptable subset of interesting data, creating actionable information to continually track and identify patterns or conditions that typically cause or lead to the failure events. The structured data set can then be queried, displayed, alerted against and correlated to user jobs running on the system. Table I shows the the approach taken by the LLSC which focuses on system problem identification and then mines the data to provide the necessary information to identify when a problem occurs. The last crucial feature of the collected data set is flexibility. The ability to change and expand the data included and adjust the alert threshold levels is an integral part of the architecture used in developing the LLSC data set.

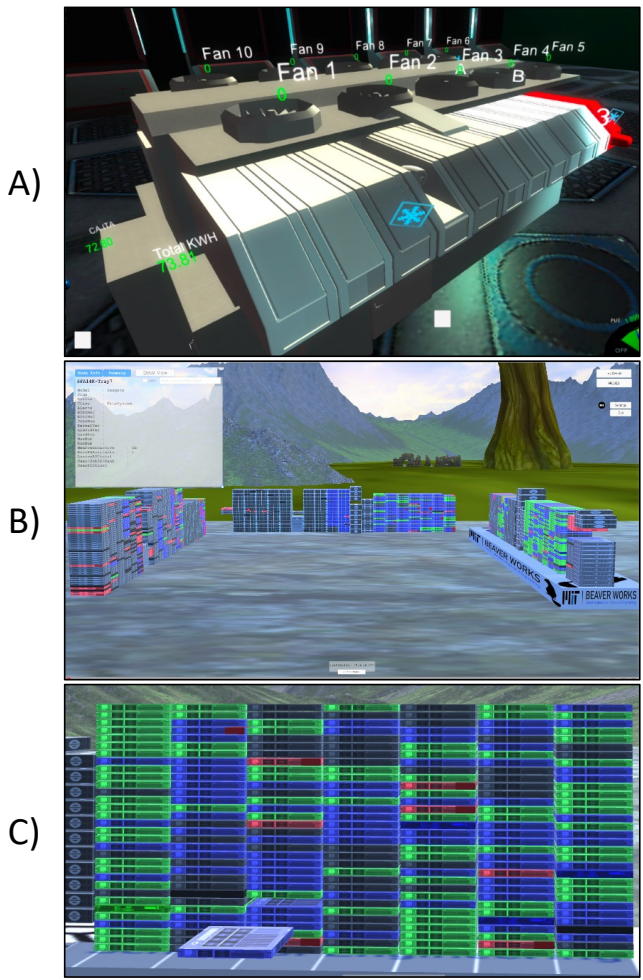

TABLE I: *Problem identification by traditional means vs 3D graphical representation in real time.*

| Node State | Symptom | Traditional | MM3D |
|---|---|---|---|
| Offline | System unreachable | Ping / vendor BMC | Whole node turns red |
| Hardware problem | Transient machine failures | Vendor BMC / log farm | Component highlighted |
| Out of sync with system | Failed job execution | Package validation / test | System version alert |
| Low Memory | Memory locked / unavailable | Escalated privileged tools | Memory visual alert |
| CPU Load | System sluggish | Escalated privileged tools | Fan speeds / alert light |
| GPU Load | Resources not avail | Escalated privileged tools | Fan speeds / alert light |
| Storage Available | Local data write fail | Escalated privileged tools | Disk usage lights |
| User Jobs | User experience failure | Scheduler logs access | Visual texture on node |
| Temperature | System reboot / job fails | Vendor BMC / Datacenter | Discrete system temp |
| Power usage | Cpu Performance slows | Vendor BMC / PDU access | Fan speed visual / value |
| Mounted File system | Data not available / job fail | Escalated privileged tools | Visual alert light |

*B. HPC Analytics*

Once the data is restructured into useful information the inherent processing power present in a supercomputing environment is harnessed. The LLSC HPC platform leverages the strategies and techniques commonly used in Big Data communities to store, query, analyze, and visualize voluminous amounts of data. The pipeline consists of the Apache Accumulo [14] database, Matlab [15]/Octave [16] analysis environment, and D4M. This software suite is part of the larger MIT SuperCloud environment, which has spurred the development of a number of cross-ecosystem innovations in high performance databases [17], [18], database management [19], data protection [20], database federation [4], [3], data analytics [21], and dynamic virtual machines [22], [23].

Fig. 2: *In-Game Views of EcoPod (A), Node layout (B), Rack View (C) provide situational awareness of HPC components in real time.*

from the massive amount of raw data points that are collected, and secondly, to make use of the computational resources of the HPC system itself to pre-process this information and identify the component and user statuses and alert conditions in a way that enhances situational awareness. This situational awareness is achieved by utilizing a 3D gaming environment to create a physically intuitive representation of the overall system and sub-components (see Figure 2).

*A. Data to Information*

As indicated previously, the level of available data from the system components is more than adequate to understand the way the system is functioning in real time and determine system failures or bottlenecks. The challenge is to take the large volume of unstructured time-series and log data and forge it into a working data set that captures the critical aspects of the data and makes them intuitive to understand. This is done by reversing the manual process of the experts mining the data used in search of the reasons for poor system performance or job failures. Then specifying which information is necessary

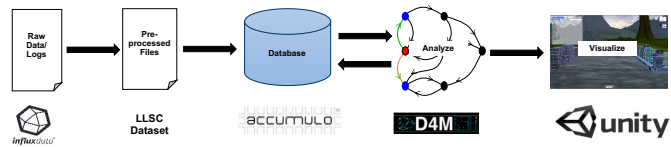

Fig. 3: *Data Flow from raw data gathering, to ordered data set, to processed data, and to 3D graphical display.*

The LLSC systems have continued to grow in scale and capabilities. The LLSC added a second Hewlett Packard Enterprise EcoPOD providing an additional dense 40 rack positions for HPC resources. The performance optimized data center is now home to our most recent system additions including TX-Green2, an Intel Xeon Platinum 8260 system consisting of 900 Nodes and over 43,000 cores. TX-Green2 debuted on the June 2016 Top500 list at number 279 in the world and is an important computational refresh to our general purpose

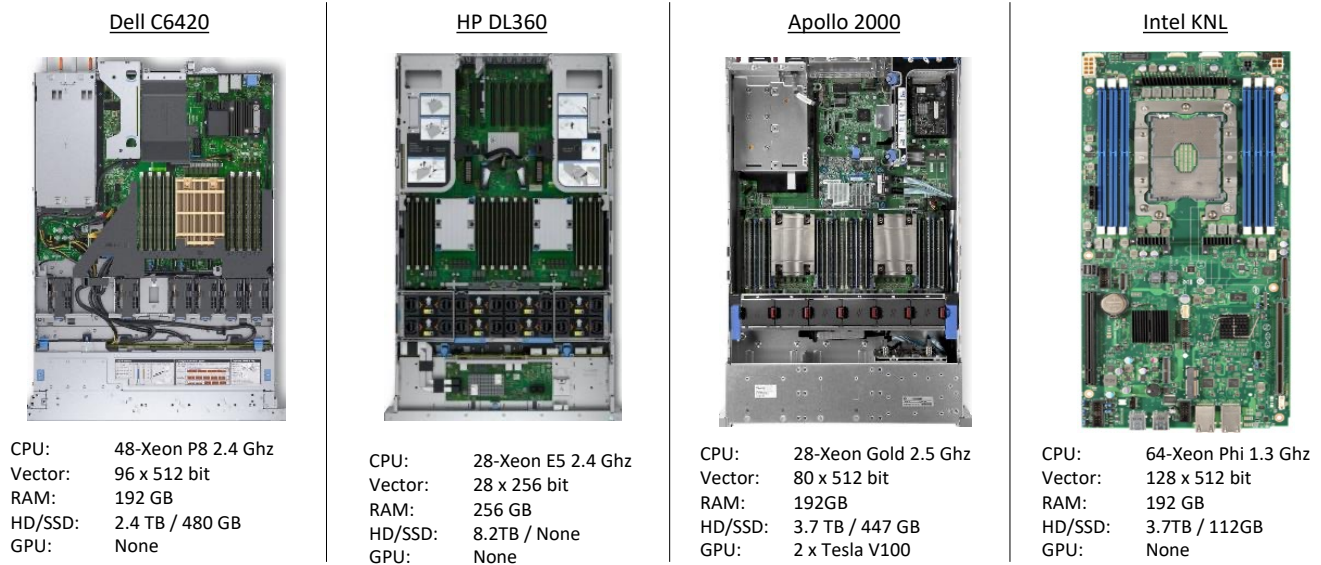

Fig. 4: *Visual representations of actual compute hardware visually convey very little information on the configuration and system load. System capacity and node characteristics like high ram, storage, compute cores, or GPU are not visually intuitive and are left for the text specifications.*

HPC user community. In 2019 the LLSC brought online TX-GAIA, which was featured as the number 42 most powerful system in the world November 2019 Top500 list [24]. The TX-GAIA system is comprised of 896 nVidia V100 GPU's and achieved 5.16 Petaflops running HPL Linpack benchmark. The TX-GAIA system is the computational backbone of MIT Lincoln Laboratory's AI and Machine Learning research. The LLSC total assets have grown to over 120,000 processing cores across two data centers backed by over 20 Petabytes of Lustre storage.

*C. 3D Gaming Visualization and Usage*

The idiom of "a picture is worth a thousand words" has been shown reasonably true in experiments where images are processed 6x-600x faster than words [25] and subjects performed significantly better using 3D displays [26]. Furthermore, the human brain can process entire images that the eye sees in as little as 13 milliseconds [27]. We live in a 3D world and our eyes and brain may have evolved to process information in this manner. These observations have led the LLSC to develop the MM3D visualization tools using the Unity 3D gaming engine. The concept of using games in work environments is not new; Clark Abt advocated the use of board games for such uses in his 1970 book Serious Games [28]. This work proposed the use of games for training, education, and business purposes and that the application to serious concepts in games should also not lose the entertainment value.

MM3D has been successful for multiple reasons. The Unity 3D game platform is widely accepted and used by millions of "gamers" who have identified the platform as an exceptional interface to convey interactive actionable data. 'Gaming environments have shown to be a far more engaging and rich vehicle to convey information than traditional web platforms [1]. Training tends to be far less of an issue using 3D gaming environments as many people, particularly those who are involved with computing, are very familiar with 3D environments for personal entertainment. This is a major advantage as cognitive absorption is an underlying determinant of the perceived usefulness and perceived ease of use [29]. The 3D gaming environment is uniquely adapted to display vast amounts of information unlike other visual mediums. Studies on 2D vs 3D interfaces indicate more natural ways to visualize hierarchical data should be strongly considered during the interface design process [26].

*D. Compute Node Visualization in 3D*

A first step in creating using a 3D environment is the depiction of the compute hardware. A simple picture of the the internal hardware shows very little differentiation between essential components (see Figure 4). Seasoned HPC administrators would have to study such a view to understand the component configuration and would have to fall back on their knowledge of the system's specifications. Because of this, we chose to deconstruct and rethink how a compute node and its internal components are displayed and how such a display would allow the user facing support team to help educate the new or less experienced user towards a better understanding of the relationship between their processing workflow and the underlying components of a supercomputer. This allows the user to better take advantage of node architecture and achieve better results and improves the overall system performance.

The 3D environment enables us to deconstruct the compute node in novel ways to explicitly identify the critical components and how they differ across architectures as seen in Figure

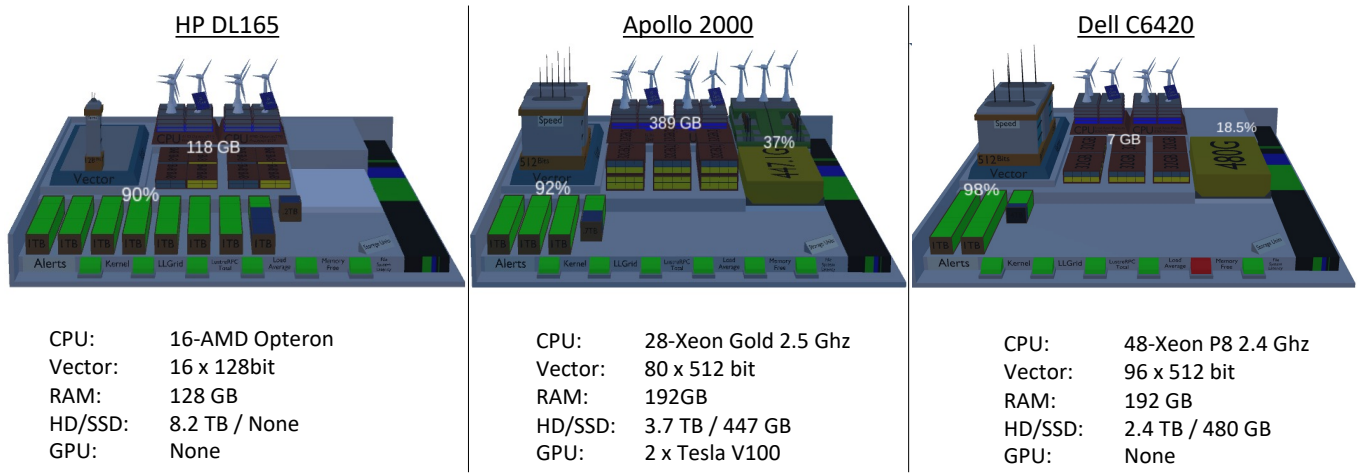

| | HP DL165 | | Apollo 2000 | | Dell C6420 |
|---|---|---|---|---|---|
| CPU: | 16-AMD Opteron | CPU: | 28-Xeon Gold 2.5 Ghz | CPU: | 48-Xeon P8 2.4 Ghz |
| Vector: | 16 x 128bit | Vector: | 80 x 512 bit | Vector: | 96 x 512 bit |
| RAM: | 128 GB | RAM: | 192GB | RAM: | 192 GB |
| HD/SSD: | 8.2 TB / None | HD/SSD: | 3.7 TB / 447 GB | HD/SSD: | 2.4 TB / 480 GB |
| GPU: | None | GPU: | 2 x Tesla V100 | GPU: | None |

Fig. 5: *Virtual Hardware representations in 3D using real time data and gaming interaction allow for higher levels of situational awareness. Virtual representations of the node hardware allow for emphasis of pertinent components, like CPU, storage, RAM, and GPU, including load and usage as well as alerting conditions. The representation can also be used as a tool to show users how their work fits on certain node types and bottlenecks on others.*

5. We are able to take advantage of the flexibility granted in the 3D world to use logical heuristics to identify CPUs, GPUs, AVX Vector units, Memory, CPU load, Disk space, and system usage. This allows for an intuitive understanding of the node usage and facilitates discussions with users on how code takes advantage of different components and how bottlenecks be avoided. Systems today have become more specialized to perform well against certain workloads, with over five different system architectures across our system the deconstructed node provides a reference point for identifying the appropriate node architecture on which to schedule the users job. This can help ensure that resources are allocated efficiently and are appropriately sized to meet the researchers needs.

## IV. Results

### A. Situational Awareness

MM3D developed over the last decade by the LLSC has proven a very effective tool for maintaining situational awareness in a complex environment. The scale and diversity of the HPC assets managed by the monitoring system has increased tenfold, notable additions including a system for MIT Campus, a second Hewlett Packard Enterprise EcoPOD, and the inclusion of assets located in the Massachusetts Green High Performance Computing Center (MGHPCC) without having to alter the basic game architecture to adjust for the increase in scale. The data volume has grown at a similar pace and both the backend ingestion and processing as well as the Unity 3D environment has been able to absorb the additional systems components with only superficial adjustments. Much like many commercial games built on Unity 3D platform we have taken advantage of the flexibility provided in 3D space and the ability to expand into larger and more varied virtual worlds.

The state of current operating conditions and component status is represented in real-time in a familiar way that the administrator or facilities operator can easily interpret. This is where the gaming environment allows you to augment the objects displayed and show an enhanced representation of the real world in a way that that more naturally cue the operators' senses to convey alerts or critical information. Using colors, shapes, and animating objects a user can quickly interpret their meaning as they would if they saw it in the real world, and this is further augmented using accepted gaming interface overlays of text or icon information which are at the margins of the display. Again, all very familiar to anyone who has played video games. Figure 2C shows the representations of a few racks of HPC nodes in the game that the user can navigate to in-game with colors indicating load, alerts, or out of service as well as the information overlays displaying information about the nodes.

The primary innovation of the current implementation of the LLSC 3D monitoring system is the use of representative models of the compute nodes to display node status and load. This strategy was employed to exploit the ability to use representative 3D graphics to more accurately visualize the pertinent information known about the hardware. This is a significant advantage since realistic views of compute hardware tell very little about the status or capacity of the component. Figure 4 shows various hardware types used by the LLSC. While experts could distinguish some capacity and configuration information with close inspection without the accompanied text it would convey very little information of situational awareness.

To compensate for this we take the approach of found in other complex environments and use a representative model. Similar to how anatomical models are often shown color coded and in a more visually striking manner to better convey pertinent biological information. Using this approach solved the problem of the limited amount of information an actual hardware picture provides and allowed building representative characterizations of the compute nodes that highlighted the subcomponents with pertinent information. Utilizing the interactive and intuitive functionality of the gaming environment, a user can drill down into various aspects of the system components and further assess the hardware, load and behavior of users on the system.

*B. Node Representation Development*

The 3D node representations used in the game were developed to portray the individual characteristics of each node type on the system and animated to represent current status across system components. The node models were designed initially through drafts in the visual arts app Procreate [30], then once they reached an agreed representation they were used as the basis for creating 3D models in the program Blender 3D [31]. Once created in a 3D space the models were iterated on and eventually animations were created in both Unity 3D and Blender 3D. These animations are capable of representing the use of sections such as the memory and storage using bright textures to create an obvious indication of change. The designs of the nodes are based on real world architecture in order to capitalize on a common understanding of buildings and associated functions. The Graphics Processing Unit for example has pipes and animated pulleys to indicate that it is in use (see Figure 6). The animated features of the node are also used to convey information; turbines featured on both the CPU and GPU can spin in congruence with data corresponding to clock rate. Prominently featured alert lights on the front of nodes animate according to the risk level of the data in important corresponding fields. The colors used across the model are intentionally low intensity and made up of light values to clarify hue without distracting visual hotspots, helping to to call attention to the aforementioned alerts and other changes that may animate elsewhere. The color palette (neglecting animated colors) is pastel-like and utilizes bright hues to encourage users to develop color relations to different sections and associated functions. This will allow users to more easily categorize the functions of a computer to promote a stronger understanding of the workings of computers, as well as supporting intuitive accessible system monitoring and awareness.

*C. Conclusion and Future Work*

Leveraging HPC Big Data analytic techniques and the 3D gaming platform for a converged DCIM solution have been explored. The resulting MM3D DCIM tool provides stakeholders with real time insight into critical facilities systems and IT infrastructure [32]. While the current implementation of the monitoring system provides excellent situational awareness there is still many paths that can be explored for further development. The backend data gathering is in the beginnings of a re-architecture to scale for the next order of magnitude growth of the current and future systems. An incorporation of additional central storage statistics, job scheduler logs, network logs, and user data into the processed data-set is underway. This will allow for enhanced methods of triggering alerts based on user behavior and their effects on system components. The goal is to provide information for HPC System personal to conduct drill-down troubleshooting and push information to user support personnel to identify non-optimized user behavior and act on this without having to engage the systems personnel. The 3D gaming environment opens up a new world in which to visualize this information and provide views to all levels of people who interface or need information from the supercomputing environment.

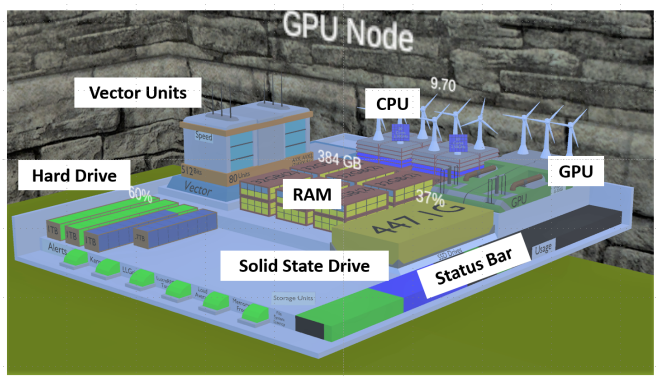

Fig. 6: *Detailed representative view of HPC compute node. Graphical representation of salient components with text labels, color, and animations used to show system load.*

ACKNOWLEDGMENTS

The authors wish to acknowledge the following individuals for their contributions and support: Bob Bond, Alan Edelman, Nathan Frey, Jeff Gottschalk, Chris Hill, Hayden Jananthan, Charles Leiserson, Dave Martinez, Joseph McDonald, Steve Rejto, Matthew Weiss, Marc Zissman.